\documentclass[12pt]{article}
\usepackage{amsmath}
\usepackage{multicol}
\usepackage[dvips]{graphicx}
\usepackage{times}
\usepackage{graphicx}

\newcommand{\bb}{\bibitem}
\newcommand{\BF}{\begin{figure}\begin{center}}
\newcommand{\EF}{\end{center}\end{figure}}
\newcommand{\BE}{\begin{equation}}
\newcommand{\EE}{\end{equation}}
\newcommand{\BEA}{\begin{eqnarray}}
\newcommand{\EEA}{\end{eqnarray}}

\begin{document}

\title{Probing the Origin of the Large-angle CMB Anomalies}
\author{Kaiki Taro Inoue}


\maketitle
\begin{abstract}
It has been argued that the large-angle cosmic
microwave background anisotropy has anomalies 
at $\sim 3\sigma$ level. We review various proposed 
ideas to explain the origin of the anomalies and
discuss how we can constrain the proposed models using
future observational data.

\end{abstract}

\section{Introduction}
There has been mounting evidence that 
the large-angle cosmic microwave background (CMB) anisotropy 
has anomalies roughly at $3\sigma$ level. 
In addition to the anomalously low 
quadrupole reported by COBE-DMR (Smoot et al. 1992),
various types of anomalies have been reported after the 
release the WMAP data (Bennett et al. 2003), namely, 
the octopole planarity and the alignment between the quadrupole
and the octopole (Tegmark et al. 2004, de Oliveira-Costa et al. 2004); 
an anomalously cold spot on angular scales 
$\sim 10^\circ$ (Vielva et al. 2004, Cruz et al. 2005);
and an asymmetry in the large-angle power
between opposite hemispheres (Eriksen et al. 2004, 
Hansen et al. 2004). Evidence for other forms of 
non-Gaussianity on large angular scales has also 
been reported (Chiang et al. 2004, Copi et al. 2004, Park 2004, 
Schwarz et al. 2004, Larson \& Wandelt 2004).

As the origin of the anomalies, a variety of solutions
have been suggested.  To explain the low quadrupole, Luminet (2003) et al. 
proposed a non-trivial spherical topology and Gordon (2004) 
proposed isocurvature perturbations due to the dark energy.
Jaffe (2005) et al. considered a locally anisotropic 
model based on the Bianchi type $\textrm{VII}_h$ universe 
to explain the quadrupole/octopole planarity and the alignment. 
Other papers have studied the possibilities that the large-angle CMB
is affected by local non-linear inhomogeneities (Moffat 2005; 
Tomita 2005a,b; Vale 2005; Cooray \& Seto 2005; Rakic et al. 2006,
Inoue \& Silk 2006a,b). In this paper, we analyze the plausibility
of these models and discuss what could be the most plausible 
scenario and how it will be probed by future observation.

\section{Large angle anomalies} 
We summarize the feature of observed anomalies on large angular scales.
Using the WMAP 1st year data (Bennett et al. 2003), 
noticeable deviations from the prediction of the 
fiducial $\Lambda$CDM cosmology (flat-FRW) are found in the angular
power spectrum at $l=2$, $l \sim 20$, and $l\sim 40$. These anomalies can be
called the statistically isotropic anomalies (SIA).
Other types of anomalies can be called statistically 
anisotropic anomalies (SAA), which include 
the octopole planarity and the alignment between the quadrupole
and the octopole (Tegmark et al. 2003, de Oliveira-Costa et al. 2004),
asymmetry in the large-angle power
between opposite hemispheres (Eriksen et al. 2004, 
Hansen et al. 2004), an anomalously cold spot on angular scales 
$\sim 10^\circ$ (Vielva et al. 2004, Cruz et al. 2005, 2006).
A mysterious correlation between the quadrupole plus octopole and the 
ecliptic plane or equinox has also been found (Copi et al. 2004).
These features are anomalous roughly at $3-\sigma$ level if
the fiducial standard LCDM model ($\Omega_0=0.24, \Omega_\Lambda=0.76$) 
is assumed.

\section{Feature on horizon scale in FRW models}
The simplest way to suppress the large-angle 
fluctuation is to consider a specific feature 
on the present horizon scale.  Such a suppression 
can be realized by introducing a cutoff on the 
primordial power spectrum. Then the ordinary Sachs-Wolfe (OSW) contribution
at the last scattering would be significantly
affected by such a cutoff while the integrated Sachs-Wolfe
 (ISW) contribution remains intact. A similar mechanism can work
if one introduces an additional isocurvature mode
that unticorrelates with the adiabatic mode. A certain class 
of the dark energy models that realize such a mechanism has been proposed
(Gordon \& Hue 2004).  Modification of dynamics of 
the curvature perturbation due to non-trivial dynamics of the 
scalar field can also suppress the ISW contribution. 
Although the quadrupole can be lowered by these
mechanism, planarity/alignment feature 
seems difficult to realize since the background metric
is spatially homogeneous and isotropic. 

\section{Non-trivial topology with finite volume}
After the discovery of the low quadrupole using the COBE data, 
suppression of the quadrupole for a toroidal topology 
$T^3$ has been studied (Starobinskly 1993, Stevens et al. 1993). 
Assuming the standard ``slow-roll''
inflationary scenario, the periodic boundary condition
on the present horizon scale due to the toroidal topology 
naturally introduces a cut off scale on the primordial 
power spectrum. Furthermore, discreteness of the 
mode function yields an oscillating feature in the power
spectrum. Although the fit to the observed angular power spectrum 
becomes better in comparison with the infinite flat FRW model,
the fit to the observed fluctuations using full covariance 
matrix defined in pixels on the sky becomes worse when normalized
over the orientation of the observer. 
However, for particular choices of orientation of the 
observer, better fits can be obtained in comparison with 
the corresponding infinite model (Inoue \& Sugiyama 2003). In other words, almost all
orientations of the observer are ruled out. 
A simple analysis using only the 
angular power spectrum can lead to a somewhat 
misleading result. 
  
After the release of the WMAP data, Luminet et al. (2003) 
considered a globally homogeneous spherical model 
with a fundamental domain described by 
a dodecahedra. For a density parameter $\Omega_0=1.013$,  
the comoving volume of the space is just $83 \%$ of that
within the last scattering surface. 
Therefore, the large scale fluctuations beyond the present horizon
can be suppressed. The low quadrupole can be obtained by such
a cut off beyond the present horizon. 
Unfortunately, this model has been ruled out by the 
``circle-in-the-sky'' analysis using the WMAP data (Cornish et al. 2003).
Furthermore, a subsequent analysis showed that the alignment/planarity 
feature in the $l=2,3$ modes cannot be naturally obtained (Weeks,
2006). 
Therefore, it seems reasonable to conclude that the non-trivial 
topology alone cannot explain all the features of the 
large-angle anomalies (SIA\&SAA).

\section{Homogeneous anisotropic models}

Jaffe et al. (2005) considered 
a certain type of locally anisotropic model called Bianchi 
$\textrm{VII}_h$ model. 
In contrast to the FRW models, there is a shear and a vorticity 
which come from the anisotropic background metric. They can account
for fluctuations with a particular
``axis'' and the ``cold spot'' in the sky.

The observational feature in Bianchi models has been studied (Barrow 1985)
and their cosmological constraint using the COBE data 
has been explored (Bunn et al. 1996 and Kogut et al. 1997).
The model has succeeded in explaining the planarity of the
$l=2,3$ and $l=5,6$ modes and the large-scale ``north-south'' 
power asymmetry and the cold spot by introducing a particular 
spiral pattern on the background gaussian fluctuations.  
However, on smaller angular scales, the fit to the data
becomes significantly worse because it needs a negatively curved 
universe with a density parameter $\Omega_0=0.5$ which is 
more than twice the ``fiducial value'' $\Omega_0=0.24$.   

Because the Bianchi $\textrm{VII}_h$ 
model necessarily introduces ''additive''
contribution to the intrinsic anisotropy, the low quadruple 
can be only achieved by unusual cancellation between them.  
Although a quantitative analysis has not been done,
this might be another problem. 
Furthermore, it cannot explain the 
correlation with the ecliptic plane or the CMB dipole because there
is no direct connection between the background geometry and the solar system.

Note that cosmological 
perturbation on this model has not been fully treated in this analysis
as the three-dimensional vector and tensor modes generally  
couple to the three-dimensional scalar mode in anisotropic model.
Simple addition of scalar perturbation on the FRW background and that 
from the anisotropic geometry may lead to a wrong result if
coupling between the scalar type perturbation and the anisotropic metric
perturbation on the present horizon scale is not negligible.

\section{Local Inhomogeneity}

As the origin of the anomalies, local inhomogeneities 
have been considered by several authors (Moffat 2005; 
Tomita 2005a,b; Vale 2005; Cooray \& Seto 2005; Raki\'c et al. 2006). 
However, none of these explanations has succeeded in explaining the 
specific features of the anomalies, namely, the octopole planarity, the alignment
between the quadrupole $(l=2)$ and 
the octopole ($l=3$), and the alignment 
between the multipoles $(l=2+3)$ with the 
ecliptic plane. For instance, if one 
applies a model in which the local group is falling into 
the center of the Shapley supercluster (SSC), 
the discrepancy between the model prediction and the 
observed data becomes even worse (Raki\'c et al. 2006). 

Inoue \& Silk (2006a,b) firstly explored the possibility 
that the CMB is affected by a small number of compensated local
dust-filled voids.  It is found that a pair of voids
with radius $(2-3)\times 10^2~h^{-1}$ Mpc and matter density contrast 
$\delta_m=-0.3$ separated by $60^\circ$ can account  for the alignment 
features in multipoles with $l=2,3,4$ and the planar features
in multipoles with  $l=2,3,6$. The Shapley supercluster (SCC)
is near the tangential point of the two local large voids. 
The mysterious correlation with the ecliptic plane 
can be explained naturally because the ecliptic plane is by chance 
tangential to the CMB dipole that originates from a mass concentration
around the SCC.  The cold spot in the Galactic southern 
hemisphere, anomalous at roughly 
$3\sigma$ level, can be also explained by such a large 
void in the direction of the cold spot at $z< 1$.    

If such dust-filled large voids exist, we would be able to observe 
dispersion in the locally measured  Hubble constant 
as measured both in different directions and at different
redshifts. For voids with a matter density contrast 
$\delta_m=-0.3$, the expected fluctuation in the Hubble
constant is as large as $2-4 \%$. 

The inflow pattern in the void wall may induce a small polarization
signal, as will the associated gravitational lensing of the CMB.
These effects are small, amounting to an imprint on the ambient
polarization pattern of order a percent, but the phase
structure would be unique and correlated with both the temperature map
and the large-scale galaxy distribution.

 \section{Future Prospects}
In order to determine the origin of the anomalies, it is of 
great importance to compare the large-scale structure (LSS) 
with the CMB anisotropy. Although the current 
signal of the LSS-CMB cross correlation is  
$\sim 3\sigma$ level at most, the precision can be greatly improved 
by future wide and deep field galaxy surveys.  
Future observation of 21cm emission from neutral hydrogen 
gas will also provide us new information about the 
three-dimensional distribution of baryons. 
Much stringent constraints on the non-trivial topology 
or Bianchi models can be obtained from such data. Measurement of 
the Hubble anisotropy using the SNIa data will be a
good test to assess to what extent the local 
inhomogeneities affects the large-angle CMB fluctuations.

\end{document}